\documentclass[twocolumn,showpacs,floats,aps,prb,groupedaddress,showpacs,amsfonts,amssymb]{revtex4}
\usepackage{bm}
\usepackage{hyphenat}
\usepackage{graphicx}
\usepackage{psfrag}
\usepackage[latin1]{inputenc}
\newlength{\sepmod}
\newcommand{\ds}{\displaystyle}
\newcommand{\ts}{\textstyle}

\def\l{\langle}
\def\L{\left\langle}
\def\r{\rangle}
\def\R{\right\rangle}

\def\mcc{\multicolumn{1}{c}}
\def\d{\mathrm{d}}

\begin{document}
\title{The Harris-Luck criterion for random lattices}

\author{Wolfhard Janke}
\email{janke@itp.uni-leipzig.de}

\author{Martin Weigel}
\email{weigel@itp.uni-leipzig.de}

\affiliation{Institut f\"ur Theoretische Physik,
  Universit\"at Leipzig, Augustusplatz 10/11, 04109 Leipzig, Germany}

\date{\today}

\begin{abstract}
  The Harris-Luck criterion judges the relevance of (potentially) spatially
  correlated, quenched disorder induced by, e.g., random bonds, randomly diluted
  sites or a quasi-periodicity of the lattice, for altering the critical behavior of
  a coupled matter system. We investigate the applicability of this type of criterion
  to the case of spin variables coupled to {\em random lattices\/}. Their aptitude to
  alter critical behavior depends on the degree of spatial correlations present,
  which is quantified by a {\em wandering exponent\/}. We consider the cases of
  Poissonian random graphs resulting from the Vorono\"\i-Delaunay construction and of
  planar, ``fat'' $\phi^3$ Feynman diagrams and precisely determine their wandering
  exponents. The resulting predictions are compared to various exact and numerical
  results for the Potts model coupled to these quenched ensembles of random graphs.
\end{abstract}


\maketitle

\section{Introduction}
\label{sec:intro}

The concept of quenched disorder coupling to the local energy density has evolved to
be the main framework for the modeling of the types of randomness found in real
physical systems \cite{cardy:book,young:book}. In the presence of frustration this
includes the vast field of spin glasses, which in the past decades has attracted an
enormous amount of analytical and numerical research, see, e.g., Refs.\ 
\onlinecite{mezard:book,fischer:book,young:book,binder:86a}. Here, we will be concerned
with the simpler case of models with purely ferromagnetic couplings. The first
investigations of this problem have considered defects distributed in the system
completely at random \cite{elliott:60a,mccoy:68a}, realized in lattice models, e.g.,
as random variation of the coupling strengths or random deletion of bonds or lattice
sites \cite{harris:74a,shalaev:94a,berche:03a}.

The relevance of this random-bond or dilution type of disorder for the universal
behavior of spin systems has been the subject of much research
\cite{harris:74a,chayes:86a,chayes:89a,ludwig:87a,ludwig:87b,ludwig:90a,luck:93a,cardy:99a}.
For the case of models undergoing a continuous phase transition on regular lattices,
Harris \cite{harris:74a} argued that for models with a specific-heat exponent
$\alpha<0$ the fluctuations in the local transition temperature induced by the
disorder degrees of freedom are not strong enough to alter universal features of the
model such as the critical exponents. Albeit not originally claimed by Harris
\cite{harris:74a}, for the converse case of a positive specific-heat exponent
$\alpha$ a significant change of the system's behavior was expected. The precise
effect of such a relevant perturbation has been the subject of some debate
\cite{harris:74a,cardy:book}. While it was originally believed that the transition
temperature fluctuations might smoothen out the phase transition so far as to
completely destroy it, it was later on realized that this, in fact, does not happen
and, instead, the system experiences a cross-over from the pure fixed point to a new,
disorder fixed point, resulting in a new set of critical exponents and further
universal properties such as amplitude ratios
\cite{ludwig:87a,ludwig:87b,shalaev:94a,ballesteros:98a,berche:02a,hellmund:02a}.
This scenario has been especially thoroughly analyzed for the case of the $q=2$, $3$,
$4$ Potts models in two dimensions, where results from perturbative methods
\cite{ludwig:87a,ludwig:87b,ludwig:90a} agree well with the outcome of numerical
simulation studies \cite{cardy:97a,jacobsen:00a,chatelain:01b,berche:03a}. From this
general observation of a smoothening effect of disorder on phase transitions, one
might expect that for systems exhibiting a first-order transition in the regular
case, disorder of the random-bond type might soften the transition to a continuous
one \cite{cardy:99a}. For the case of two dimensions it could be rigorously
established that even an infinitesimally small amount of disorder suffices to indeed
induce this behavior \cite{imry:79a,aizenman:89a,hui:89a,hui:89b}; in three
dimensions, sufficiently strong disorder coupling to the local energy density is
numerically found to soften first-order phase transitions to second-order ones, see,
e.g., Refs.\ \onlinecite{ballesteros:00a,chatelain:01a,hellmund:03a}.

Obviously, for many physical systems the assumption of an uncorrelated, isotropic
distribution of defects is not an adequate description. Instead, due to various
reasons the distribution of defects is spatially correlated. This effect occurs
isotropically due to long-range interactions between the non-magnetic ions, or in the
form of line or higher-dimensional defects
\cite{weinrib:83a,prudnikov:99a,muzy:02a,holovatch:03a,vojta:03a}. In these cases,
the reasoning leading to Harris' relevance criterion is no longer directly
applicable, but can be generalized accordingly. For the case of algebraically
decaying correlations one finds a relevance threshold depending on the dimension of
the defects as well as the power of the decay of the correlations
\cite{weinrib:83a,prudnikov:99a}. For a different model, not covered by the
random-bond paradigm, namely the co-ordination number non-periodicity found in
quasi-crystals and other aperiodic structures, Luck \cite{luck:93a} formulated a
relevance criterion that includes the situations discussed as special cases. There,
the ``break-even point'' for the relevance of randomness in terms of the
specific-heat exponent $\alpha$ is shifted from its uncorrelated value $\alpha_c=0$
to somewhere in the region $-\infty<\alpha_c\le 1$, depending on the strength of
spatial correlations of the disorder degrees of freedom measured by a geometrical
fluctuation or {\em wandering exponent\/}. A structure conceptually related to these
aperiodic models is given by different varieties of {\em random graphs\/} such as
Poissonian Vorono\"\i-Delaunay triangulations \cite{okabe:book,schliecker:02a} or the
planar, combinatorial triangulations encountered in the dynamical triangulations
approach to quantum gravity resp.\ the dual planar and orientable $\phi^3$ Feynman
diagrams \cite{ambjorn:book}. Although spin models on quenched ensembles of these
graphs have been considered in numerical simulation studies
\cite{espriu:86a,wj:93b,wj:94a,wj:95b,anagnostopoulos:99a,wj:00a,wj:00b,wj:02b}, no
connection has been made as yet with the predictions of Luck \cite{luck:93a} for
general systems with connectivity disorder. In particular, the geometrical
fluctuation exponents appearing in this relevance criterion have not been determined
for the case of these random graphs.

The rest of this paper is organized as follows. In Section \ref{sec:graphs} we
introduce the ensembles of graphs to be considered, elaborate on their generation in
a computer experiment and review some of their known properties. Section
\ref{sec:relevance} is devoted to a presentation of a version of the Harris-Luck
relevance criterion suitable to be applied to the case of random graphs considered
here and a discussion of its connection with previous results. In Section
\ref{sec:wandering} we introduce different methods to precisely determine the
wandering exponent for both graph types and present the results obtained. Finally,
Section \ref{sec:conclusions} contains our conclusions.

\section{Random Graphs}
\label{sec:graphs}

\begin{figure}[tb]
  \centering
  \includegraphics[clip=true,keepaspectratio=true,width=7cm]{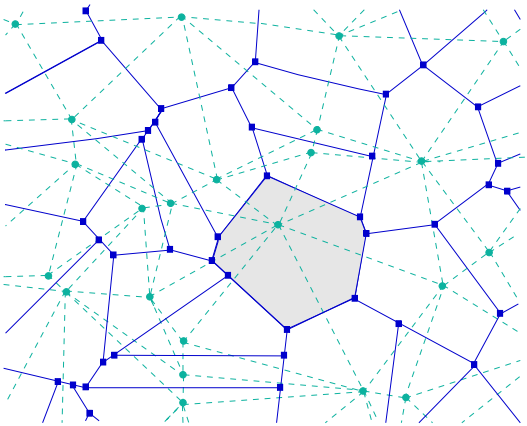}
  \caption
  {Patch of a Poissonian Vorono\"\i-Delaunay triangulation of the plane. The
    spherical, shaded dots denote the randomly distributed generators. The solid
    lines enclosing these generators define convex polygonal areas whose points are
    closer to the origin than to any other generator. They form the network of
    Vorono\"{\i} cells, which results in a three-valent graph whose vertices are
    depicted by the dark boxes.  Its geometrical dual, consisting of the generators
    and the connecting dashed lines, is the Delaunay triangulation.}
  \label{fig:voro_snap}
\end{figure}

In the following we present construction techniques and properties of two different
kinds of two-dimensional topological graphs whose randomness is solely encoded in the
degree distribution of their vertices, resulting in a topological connectivity
disorder. In contrast to the generic random graphs discussed in the context of
scale-free networks, small-world models, etc.\ \cite{albert:02a,igloi:02a}, the
ensembles of random structures considered here are {\em not\/} fully determined by
their degree or co-ordination number distributions, but additionally exhibit a
well-defined topology and long-range correlations of their disorder degrees of
freedom.

\subsection{Poissonian Vorono\"\i-Delaunay Triangulations}
\label{sec:voronoi}

\begin{figure}[tb]
  \centering
  \begin{tabular}{crrr}
    \raisebox{2.15cm}{(a)} & \includegraphics[clip=true,keepaspectratio=true,width=2.5cm]{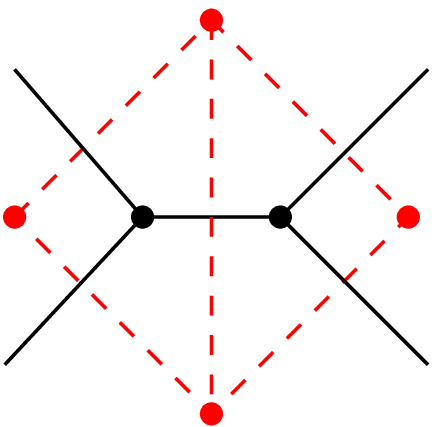} &
    \raisebox{1.12cm}{\hspace{0.25cm}\includegraphics[clip=true,keepaspectratio=true,width=1.25cm]{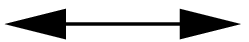}} &
    \includegraphics[clip=true,keepaspectratio=true,width=2.5cm]{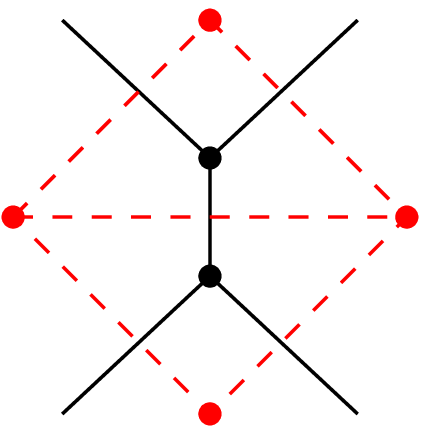}\\
    & & & \\
    \raisebox{4.4cm}{(b)} & \multicolumn{3}{c}{
      \includegraphics[clip=true,keepaspectratio=true,width=7cm]{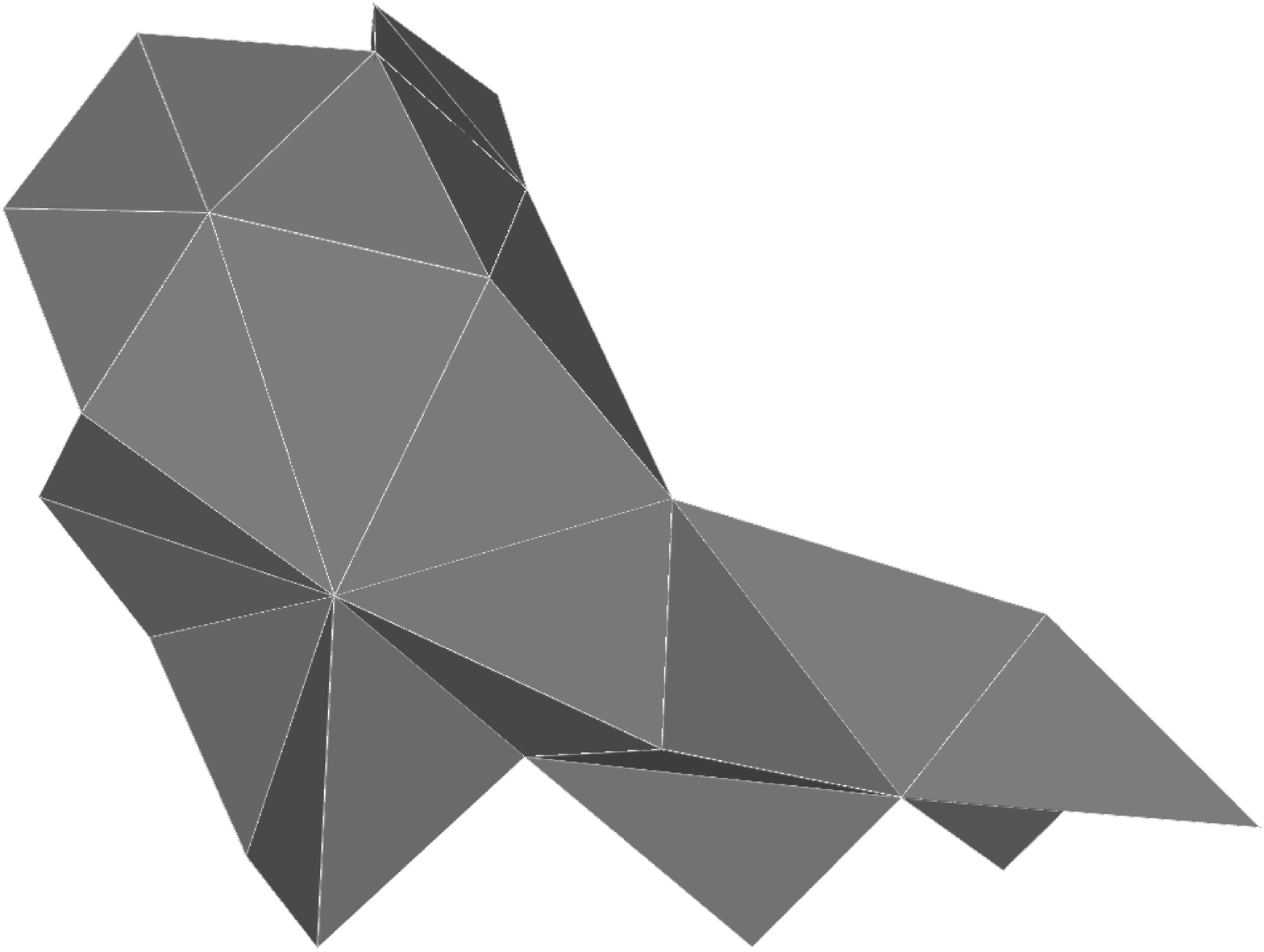}      
    }
  \end{tabular}
  \vspace*{-0.4cm}
  \caption
  {(a) The link-flip move on two adjacent triangles of a dynamical triangulation
    (dashed lines). The solid lines denote the corresponding dual, three-valent
    $\phi^3$ graphs.  (b) Example of a spherical dynamical triangulation embedded
    into three-dimensional space.  Note that the requirement of equilaterality of the
    triangles is only approximately fulfilled for the embedding shown, since there is
    no generic exact embedding algorithm for performing it.}
  \label{fig:phi3_snap}
\end{figure}

Irregular cell structures or froths appear in a large variety of natural systems,
such as foams or biological tissues \cite{schliecker:02a}. The common inverse problem
of constructing such a connectivity or cell structure from a given irregular
arrangement of vertices (so-called generators) is solved by the Vorono\"\i-Delaunay
construction \cite{okabe:book}. In two dimensions, the Vorono\"{\i} cell of a given
generator is a convex polygon around it, enclosing the part of its neighborhood which
is closer to it than to any other generator, cf.\ Fig.~\ref{fig:voro_snap}.  This is
in complete analogy with the notion of a Wigner-Seitz elementary cell in
crystallography. This construction results in the three-valent Vorono\"{\i} graph and
the dual Delaunay triangulation. If the generators are located completely at random,
as is the case in the example of Fig.~\ref{fig:voro_snap}, the resulting graph is
known as Poissonian Vorono\"\i-Delaunay triangulation \cite{okabe:book}.

These lattices are randomly disordered in several respects: edge lengths, cell
volumes etc.\ vary, as well as the co-ordination numbers $q$ of the vertices of the
Delaunay triangulation. To facilitate comparison with the second type of random
graphs to be introduced below, we here restrict ourselves to the latter aspect of
variable co-ordination numbers, i.e., we do not take any effects from variable
lengths or areas into account. Thus, the distribution $P(q)$ of co-ordination numbers
is the only random variable involved. Additionally, to eliminate surface effects, the
generators are randomly placed on the surface of a sphere instead of a patch of the
plane. Thus, the resulting random graphs are triangulations of spherical topology.

From the Euler relations, the average co\hyp{}ordination number is a topological
invariant for a fixed number of triangles in two dimensions, given by
\cite{ambjorn:book}
\begin{equation}
  \bar{q} = \frac{1}{N}\sum_i q_i = 6\frac{N}{N+4},
  \label{euler}
\end{equation}
for any closed, spherical triangulation, where $N$ denotes the number of triangles.
In the limit of infinite triangulations, $N\rightarrow\infty$, one thus obviously has
$\l\bar{q}\r=\l q_i\r = 6$. The second moment of $q$ is not exactly known, but is
numerically found to be \cite{drouffe:84a,okabe:book}
\begin{equation}
  \mu_2 \equiv \l q_i^2\r-\l q_i\r^2 \approx 1.781,
\end{equation}
as $N\rightarrow\infty$. It turns out that the random variables $q_i$ are not
independently distributed, but are reflecting a spatial correlation of the disorder
degrees of freedom in addition to the trivial correlation induced by the constraint
(\ref{euler}). The form of these correlations for nearest-neighbor vertices is
commonly described by the Aboav-Weaire law \cite{okabe:book}, which states that the
total expected number of edges of the neighbors of a $q$-sided cell, $q\,m(q)$,
should vary linearly with $q$,
\begin{equation}
q\,m(q) = (6-a)q+b,  
\label{aboav}
\end{equation}
where $a$ and $b$ are some parameters. In turns out, however, that Eq.~(\ref{aboav}),
albeit being a good effective description for a large variety of cell systems
including the case of Poissonian Vorono\"{\i}-Delaunay triangulations, is merely a
leading-order and not an exact property of these systems \cite{schliecker:02a}.

\begin{figure}[tb]
  \centering
  \includegraphics[clip=true,keepaspectratio=true,width=\columnwidth]{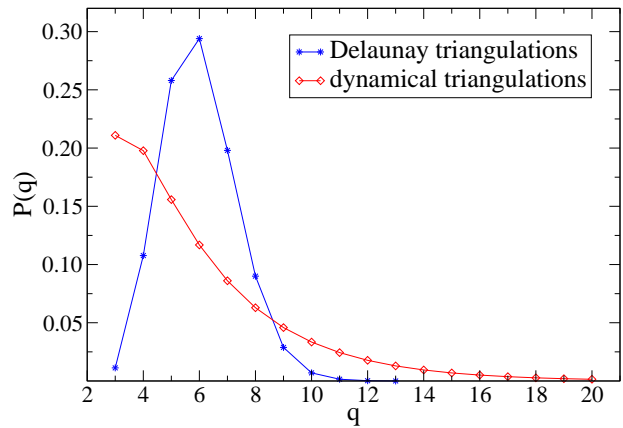}
  \caption
  {Comparison of the degree distributions $P(q)$ of Poissonian Delaunay
    triangulations and dynamical triangulations in the limit of an infinite number of
    triangles. The results are taken from Refs.\ 
    \onlinecite{drouffe:84a,boulatov:86a}.}
  \label{fig:poisson}
\end{figure}

\subsection{Dynamical Triangulations}
\label{sec:dynam}

An ensemble of planar triangulations with properties very different from those of the
Poissonian Vorono\"{\i}-Delaunay lattices is given by the so-called dynamical
triangulations model, which has been used as a constructive approach to discrete
Euclidean quantum gravity in two dimensions \cite{ambjorn:book}. This ensemble of
combinatorial triangulations is defined as that of all gluings of equilateral
triangles to closed surfaces of a given topology (such as, e.g., that of a sphere),
where all gluings are counted with equal probability. A dynamical way of generating
such graphs is the repeated application of so-called link-flip moves which re-wire a
given network [see Fig.~\ref{fig:phi3_snap}(a)]. It is known that this procedure
converges to a limiting distribution with specific properties described below. Since
this flip dynamics introduces large temporal correlations between subsequent random
triangulations, we do not actually use it to generate the graph instances needed
below, but instead revert to a recursive insertion technique known to yield the same
graph distribution, but independent graph realizations in each step
\cite{agishtein:91b}. In this construction, no edge length differences are involved
and, again, the randomness is solely encoded in the degree distribution $P(q)$ of the
vertices. Although these objects are entirely defined in terms of their intrinsic
connectivity properties, an embedding of an example dynamical triangulation into
three-dimensional space is shown for illustration purposes in
Fig.~\ref{fig:phi3_snap}(b). Technically, the geometrical duals of these
triangulations are given by the ensemble of planar, ``fat'' (i.e., orientable)
$\phi^3$ Feynman diagrams without tadpoles and self-energy insertions.  The
statistics of these objects can be explicitly treated by means of matrix models
\cite{brezin:78a,mehta:book}, leading to exact solutions for the co-ordination number
distribution $P(q)$ and many other properties of the model
\cite{boulatov:86a,ambjorn:book}. Compared to the Poissonian Vorono\"{\i}-Delaunay
model, fluctuations are much more pronounced for these graphs, and it can be shown
that the variance of co-ordination numbers approaches
\cite{boulatov:86a,godreche:92a}
\begin{equation}
  \mu_2 = 21/2
\end{equation}
as $N\rightarrow\infty$, whereas Eq.\ (\ref{euler}) still holds. Also, the
Aboav-Weaire law (\ref{aboav}) correctly describes the leading-order behavior of the
nearest-neighbor correlations \cite{godreche:92a}.

Considering more geometric properties of the graphs, the differences between both
ensembles become even more pronounced. Especially, the quantum gravity graphs of the
dynamical triangulations model can be shown to be highly fractal, being
self-similarly composed of ``baby universes'' branching off from the main surface,
i.e., macroscopical subgraphs attached to the main body by only a few links
\cite{jain:92a}. This fractal structure leads to an exceptionally large internal
Hausdorff dimension of $d_h=4$ as compared to the topological dimension of two
\cite{kawai:93a,watabiki:95a}, whereas for the Poissonian Vorono\"{\i}-Delaunay
triangulations the Hausdorff dimension remains at the trivial value of the
topological dimension. The differences between the two graph ensembles are also very
strikingly seen in the distributions $P(q)$ of co-ordination numbers as depicted in
Fig.~\ref{fig:poisson}. While $P(q)$ is peaked at $q=6$ for the Delaunay
triangulations, it drops monotonically starting from $q=3$ for the case of dynamical
triangulations, and large co-ordination numbers are much more probable for the latter
lattice type.  It can be shown that for large co-ordination numbers the distribution
$P(q)$ falls off as $\exp(-\sigma q\ln q)$ with $\sigma\approx 2$ for Poissonian
random lattices \cite{drouffe:84a}, whereas for dynamical triangulations it declines
much slower proportional to $\exp(-\sigma q)$ with $\sigma = \ln (4/3) \approx 0.3$
(Ref.~\onlinecite{boulatov:86a}).

\section{The Relevance Criterion}
\label{sec:relevance}

Trying to decide for which kind of models disorder of the random-bond type
constitutes a relevant perturbation, Harris \cite{harris:74a} suggested the following
line of reasoning. For a system of uncorrelated random bonds, the fluctuations
$\sigma_R(J)$ of the average coupling $J$ in patches of the lattice decline with the
linear patch size $R$ according to the central limit theorem, that is,
\begin{equation}
  \sigma_R(J) \sim R^{-d/2},
  \label{eq:central_limit}
\end{equation}
where $d$ denotes the spatial dimension of the system. Since disorder of the
random-bond type couples to the local energy density, local fluctuations of the
coupling induce such fluctuations of the effective local transition temperature,
which decline identically with increasing patch size.  Approaching the critical point
$t\equiv (T-T_c)/T_c = 0$, the fluctuations in a correlation volume scale as
\begin{equation}
  \sigma_\xi(J) \sim \xi^{-d/2} \sim t^{\nu d/2},
\end{equation}
where the power-law divergence $\xi \sim t^{-\nu}$ of the correlation length was
used. For the critical behavior of the pure system to persist, these fluctuations
should die out as the reduced temperature is linearly tuned to the critical point
$t=0$, i.e., one should have $\nu d/2 > 1$ or, with appeal to hyper-scaling, $\alpha
< 0$.  In the converse case $\alpha > 0$ an altered universal behavior might me
expected.  Later on, Chayes {\em et al.\/}\cite{chayes:89a,chayes:86a} showed that
under quite general conditions the specific-heat exponent of the {\em disordered\/}
system, $\alpha_\mathrm{dis}$, should be negative. This seemed plausible, since
$\alpha$ was believed to coincide with the crossover or stability exponent of the
respective renormalization-group (RG) fixed point \cite{domany:75a,kinzel:81a}. Later
on, however, it was claimed that in some cases the stability exponent might differ
from $\alpha$, such that even for positive $\alpha$ the regular critical behavior
would prevail \cite{andelman:84a}.  Several examples of such behavior, and even the
opposite case of disorder being a relevant perturbation albeit $\alpha<0$, have been
explicitly constructed for the case of hierarchical lattices
\cite{derrida:85a,mukherji:95a,magalhaes:98a,haddad:00a,efrat:01a}. Thus, the general
validity of the relevance criterion implied by Harris' argument has recently been the
object of some debate, see, e.g., Refs.\ 
\onlinecite{pazmandi:97a,aharony:98a,korzhenevskii:98a}.

As soon as the assumption of uncorrelated disorder degrees of freedom is relaxed,
Harris' reasoning is no longer applicable as it stands. For a random-bond model with
long-range correlations of the bond variables, Weinrib and Halperin
\cite{weinrib:83a} performed a renormalization group calculation, which was later on
refined \cite{prudnikov:99a}. They find that, if the disorder degrees of freedom are
correlated algebraically with a decay exponent $a<d$, disorder is irrelevant if
$\alpha < 2-2d/a$, whereas in the opposite case, the system flows towards a new,
long-range correlated disorder fixed point. For more general disorder degrees of
freedom and types of correlations including the ensembles of random graphs considered
here, this argument can be adapted in the spirit of Luck's reasoning for the case of
aperiodic structures \cite{luck:93a} as follows. Consider a patch $P$ of spherical
shape with radius $R$ and a volume of $B(R)$ vertices\footnote{All distances on the
  graphs considered in this paper are to be understood as the unique number of links
  in the shortest path of links connecting two vertices.} on a given realization of a
triangulation. The average co\hyp{}ordination number in $P$,
\begin{equation}
   J(R) \equiv \frac{1}{B(R)}\sum_{i\in P} q_i,
  \label{mean_J}
\end{equation}
fluctuates around its expected value $J_0=\bar{q}$ [cf.\ Eq.\ (\ref{euler})]. As the
size of the patch is increased, $R\rightarrow\infty$, these fluctuations decay as
\begin{equation}
  \sigma_R(J) \equiv \l|J(R)-J_0|\r/J_0 \sim
  \l B(R)\r^{-(1-\omega)} \sim R^{-d_h(1-\omega)},
  \label{fluct}
\end{equation}
defining the wandering exponent $\omega$ of the considered graph type. The Hausdorff
dimension $d_h$ enters here to account for the cases of fractal graphs. In Eq.\ 
(\ref{fluct}), the averages $\l\cdot\r$ are to be understood as the ensemble averages
of the considered class of graphs of a given total size. While for $\omega=1/2$ the
usual $1/\sqrt{\l B(R)\r}$ behavior of uncorrelated random variables is recovered,
for random lattices with long-range correlations of the co\hyp{}ordination numbers
one expects $\omega>1/2$, leading to a slowed-down decay of fluctuations.  Near
criticality, the fluctuation $\sigma_\xi(J)$ of the average co\hyp{}ordination number
in a correlation volume induces a local shift of the transition temperature
proportional to $|t|^{d_h\nu(1-\omega)}\mu_2^{1/2}$. For the regular critical
behavior to persist, these fluctuations should die out as the critical point $t=0$ is
approached. This is the case when $\omega$ does not exceed the threshold value
\begin{equation}
  \omega_c = 1-\frac{1}{d_h\nu} = \frac{1-\alpha}{2-\alpha},
  \label{luckcrit}
\end{equation}
provided that hyper-scaling is in effect. Conversely, for fluctuations satisfying
$\omega>\omega_c$ a new type of critical behavior could occur. Re-writing Eq.\ 
(\ref{luckcrit}) as
\begin{equation}
  \alpha_c = \frac{1-2\omega}{1-\omega},
  \label{luckother}
\end{equation}
it is obvious that for $\omega=1/2$ the Harris criterion is recovered.

It is easily seen that the case of algebraically decaying correlations discussed in
Ref.\ \onlinecite{weinrib:83a} is included in Eq.\ (\ref{luckother}) as a special case.
Specifically, for the case of connectivity disorder considered here, an isotropic
power-law correlation would be given by
\begin{equation}
  G_{qq}(i,j)\equiv\l\delta q_i\delta q_j\r \sim \mathrm{dist}(i,j)^{-a},
  \label{eq:corr}
\end{equation}
where $\delta q_i = q_i-\bar{q}$ denotes the co-ordination number defect at vertex
$i$ and the distance $\mathrm{dist}(i,j)$ is defined as the unique number of links in
the shortest path of links connecting the two vertices $i$ and $j$. Then, the
fluctuation of the mean $\overline{\delta q_R}$ over a spherical patch $P$ of radius $R$
is given by
\begin{equation}
  \begin{array}{rcl}
  \sigma^2(\overline{\delta q_R}) & = & \ds \frac{\sigma^2(\delta q)}{R^{d_h}} +
  \frac{1}{R^{2d_h}}\sum_{i\neq j\in P}\l\delta q_i\delta q_j\r \\
  & \sim & \ds \mathrm{const}\times R^{-d_h} + \mathrm{const}\times R^{-a} \\
  \end{array}
\end{equation}
as $R\rightarrow\infty$. For short-range correlations $a\ge d_h$ the leading behavior
is that of Eq.~(\ref{eq:central_limit}), such that the Harris criterion stays in
effect. For long-range correlations $a<d_h$, on the other hand, the leading term is
proportional to $R^{-a}$. Comparing to Eq.\ (\ref{fluct}), we arrive at the following
expression,
\begin{equation}
  \omega=1-a/2d_h,
  \label{eq:omega_a}
\end{equation}
such that from Eq.~(\ref{luckother}) we arrive at $\alpha_c = 2-2d_h/a$, in agreement
with the direct observation in Ref.~\onlinecite{weinrib:83a}.

Since for systems with sufficiently long-range correlations of the disorder degrees
of freedom $\omega>1/2$, such correlated disorder is {\em more\/} relevant than
uncorrelated randomness in the sense that a change of universality class can already
be expected for some range of {\em negative\/} values of $\alpha$, cf.\ Eq.\ 
(\ref{luckother}). If, on the other hand, correlations decay exponentially, the
threshold $\alpha_c=0$ of the Harris criterion should stay in effect.

\section{Correlators and Wandering Exponents}
\label{sec:wandering}

In view of the discussion presented in Sec.\ \ref{sec:relevance}, two complementary
approaches towards a numerical determination of the wandering exponent $\omega$
present themselves, either a direct evaluation of the scaling of the fluctuations
defined in Eq.\ (\ref{fluct}) or an analysis of the correlation function
$G_{qq}(i,j)$ of the disorder degrees of freedom defined in Eq.\ (\ref{eq:corr}) to
infer an estimate of $\omega$ via Eq.\ (\ref{eq:omega_a}). Both of these methods will
be applied for the cases of Poissonian Vorono\"{\i}-Delaunay as well as dynamical
triangulations. For both graph ensembles, a quenched average has to be taken over a
number of graph realizations of the considered ensemble. For this purpose, we
generate lattices of spherical topology and consider the triangulations of varying
co-ordination numbers as the basic objects and refer to their geometric duals as the
``dual'' lattices.

\subsection{Analysis of Correlation Functions}
\label{sec:correlators}

\begin{figure}[tb]
  \centering
  \includegraphics[clip=true,keepaspectratio=true,width=\columnwidth]{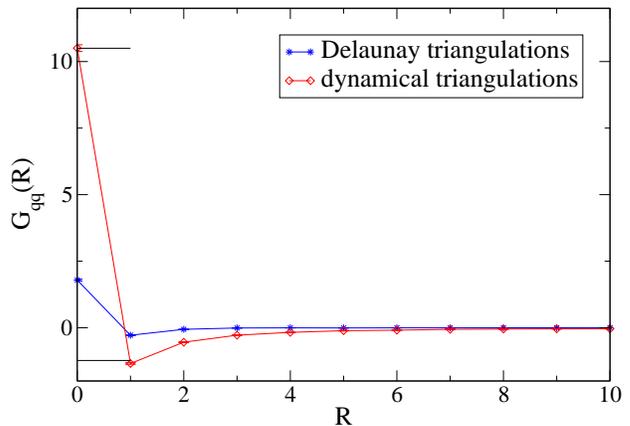}
  \caption
  {Comparison of the connected correlator $G_{qq}(R)$ of the co-ordination number
    defects $\delta q =q_i-\bar{q}$ of Poissonian Delaunay triangulations and
    dynamical triangulations of size $N=125\,000$ triangles. The connecting lines are
    only drawn to guide the eye. The two short horizontal lines indicate the exact
    result $\mu_2=10.5$ (cf.\ Ref.\ \onlinecite{godreche:92a}) and the value
    $G_{qq}(1)\approx -1.2295$ for the dynamical triangulations discussed in App.\ 
    \ref{sec:near-neighb-corr}, both valid in the limit $N\rightarrow\infty$.}
  \label{fig:corr_overview}
\end{figure}

For the numerical determination of the correlator $G_{qq}(i,j)$ a decomposition of
the graphs into spherical shells around a given vertex is performed. This is done by
first picking a vertex of the triangulation at random and a subsequent slicing of the
graph into shells of equal geodesic link-distance around that vertex. For that given
choice of initial vertex $i_0$, two fundamental observables can be measured, the
volumes of the decomposition shells,
\begin{equation}
  \label{eq:shell_volume}
  g^{i_0}_{11}(R) = \sum_{\mathrm{dist}(i,i_0)=R} 1
\end{equation}
which, if properly averaged over, give the correlator of the unit operator, and,
\begin{equation}
  \label{eq:corr_before_average}
  g^{i_0}_{qq}(R) = \sum_{\mathrm{dist}(i,i_0)=R} \delta q_{i_0} \delta q_i,
\end{equation}
the average of which gives the correlator of Eq.\ (\ref{eq:corr}). In view of Eq.\ 
(\ref{mean_J}) it is obvious that $B(R)=\sum_{r=0}^R g^{i_0}_{11}(r)$. In the context
of the dynamical triangulations approach to quantum gravity, there has been some
debate on how to properly define (connected) correlators on an ensemble of random
graphs, see, e.g., Ref.\ \onlinecite{ambjorn:99a}. The uncertainty concerns the order of
taking the averages over a single graph and the graph ensemble, which has not been
explicitly specified in Eq.\ (\ref{eq:corr}). The two possibilities are given by
averaging the expressions of Eqs.\ (\ref{eq:shell_volume}) and
(\ref{eq:corr_before_average}) individually, i.e.,
\begin{equation}
  \label{eq:individ_av}
  G_{qq}(R) = \frac{\l g^{i_0}_{qq}(R)\r}{\l g^{i_0}_{11}(R)\r},
\end{equation}
where the average $\l\cdot\r$ denotes a combined average over the starting vertices
$i_0$ and the graph realizations under consideration, which is obviously equivalent
to the ensemble average of the graphs. The additional average over the starting
vertices $i_0$ is merely included to improve the statistical accuracy in practical
applications. On the other hand, one could also average on the level of the fraction,
\begin{equation}
  \label{eq:fract_av}
  G_{qq}(R) = \L\frac{g^{i_0}_{qq}(R)}{g^{i_0}_{11}(R)}\R.
\end{equation}
It turns out that, at least for the quantum gravity graphs, these two ways of
performing the average yield strikingly different results, even in the thermodynamic
limit \cite{ambjorn:99a}. The main motivation for using Eq.\ (\ref{eq:individ_av}) is
that an expression of the form $\l g^{i_0}_{\phi\phi}(R)\r$, if integrated over all
distances $R$, still yields some kind of susceptibility of the associated operator
$\phi$. Otherwise, however, correlators defined according to Eq.\ 
(\ref{eq:individ_av}) behave rather pathologically. Definition (\ref{eq:fract_av}),
on the other hand, corresponds to the natural probabilistic definition of the average
correlation of a given quantity at distance $R$, and is thus the unique ``correct''
definition in the given context and will be used throughout. Note, that in this case
the average has to be performed carefully, since the maximum linear separation
$R_\mathrm{max}$ of two points on the graph is not universal, but depends on the
graph realization as well as on the chosen initial vertex.

\begin{figure}[tb]
  \centering
  \includegraphics[clip=true,keepaspectratio=true,width=\columnwidth]{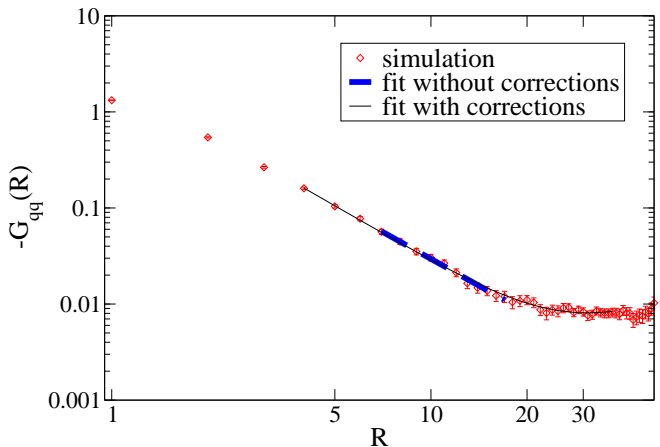}
  \caption
  {Decay of the correlator $-G_{qq}(R)$ of the co-ordination numbers of dynamical
    triangulations with $N=500\,000$ triangles in a double logarithmic plot. The
    lines show fits of the functional form (\ref{eq:decay}) to the data, where either
    the correction term was omitted ($B=0$, thick dashed line) or included with $B$
    and $b$ as free parameters (solid line). The ranges of the lines indicate the
    range of data points included in the fits.}
  \label{fig:corr_fits}
\end{figure}

Figure \ref{fig:corr_overview} shows an overview of the short-distance behavior of
the correlator $G_{qq}(R)$ defined according to Eq.\ (\ref{eq:fract_av}) for
Poissonian Delaunay as well as for dynamical triangulations as measured by averaging
over 100 graph realizations sampled with one different starting vertex per 1000 graph
vertices each. The statistical errors here and in the following were determined via
jackknifing (see, e.g., Refs.\ \onlinecite{efron:82,efron:book}) over the 100
different graph realizations. Obviously, $G_{qq}(0)=\mu_2$, corresponding to the
values cited above in Sec.\ \ref{sec:graphs}.  Note, that the negative correlation at
distances $R>0$ reflects the fact, expressed in the Aboav-Weaire law (\ref{aboav}),
that a vertex with co-ordination number $q>6$ tends to have neighbors with $q<6$ and
vice versa. The correlation for nearest-neighbor points for the case of dynamical
triangulations can be calculated in the thermodynamic limit by a series expansion of
the results found in Ref.\ \onlinecite{godreche:92a}, yielding $G_{qq}(1) \approx
-1.2295$, cf.\ App.\ \ref{sec:near-neighb-corr}. This result is in reasonable
agreement with the value found numerically for the example of $N=125\,000$ triangles
presented in Fig.\ \ref{fig:corr_overview}, $G_{qq}(1)\approx -1.34$. The remaining
difference gives a first indication of the presence of large finite-size corrections
for the case of the highly fractal quantum gravity graphs. Going beyond $R=1$, a
short glance at Fig.\ \ref{fig:corr_overview} reveals that the correlations are much
more long-ranged for the case of the quantum gravity graphs than for the
Vorono\"{\i}-Delaunay random lattices. In fact, for the Delaunay triangulations we
find that for the graph sizes up to $N=500\,000$ triangles considered, co-ordination
numbers of vertices at distances $R\gtrsim 10$ apart are effectively uncorrelated up
to the precision of our calculations. Due to this smallness of correlations it is not
possible to determine their exact functional form from the accuracy of our
measurements. However, a simple exponential decay can be fitted reasonably well to
the behavior found.  Additional checks in support of this picture will be presented
below in this Section and in Sec.\ \ref{sec:average_fluct}.

\begin{table*}[tb]
  \caption{Results of fits of the functional form (\ref{eq:decay}) to the correlator
    $G_{qq}(R)$ for dynamical triangulations with $N=500\,000$ triangles and with
    various restrictions.}
  \begin{center}
    \begin{tabular*}{\linewidth}{@{\extracolsep{\fill}}cllllcl}
      \toprule
      Restriction & \mcc{$A$} & \mcc{$a$} & \mcc{$B$} & \mcc{$b$} &
      $R_\mathrm{min}$--$R_\mathrm{max}$ & \mcc{$Q$} \\ \colrule
      $B=0$ & $-2.11(45)$ & 1.854(105) & \mcc{0} & \mcc{--} & 7--17 & 0.981 \\
      $b=3$ & $-2.37(29)$ & 1.938(92) & $5.6(19)\times 10^{-5}$ & \mcc{3}  & 4--37 & 0.999 \\
      $b=4$ & $-2.04(21)$ & 1.841(75) & $1.27(40)\times 10^{-6}$ & \mcc{4}  & 4--37 & 0.994 \\
      none  & $-2.30(24)$ & 1.919(76) & $3.1(67)\times 10^{-5}$  & 3.15(57) & 4--37 & 0.999 \\ \botrule
      \label{tab:phi3fits}
    \end{tabular*}
  \end{center}
\end{table*}

\begin{figure}[tb]
  \centering
  \includegraphics[clip=true,keepaspectratio=true,width=\columnwidth]{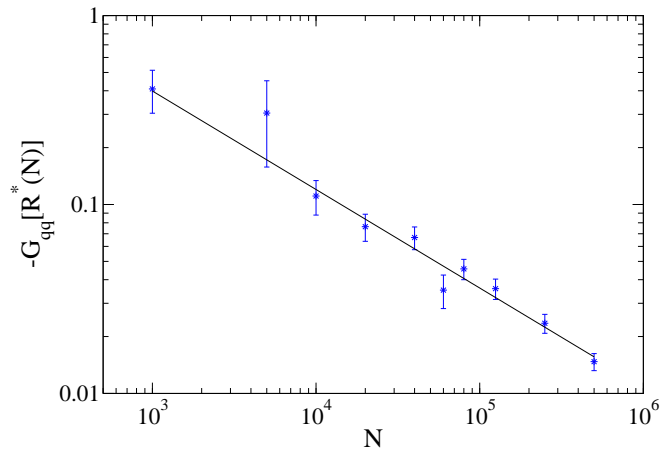}
  \caption
  {Finite-size scaling of the co-ordination number correlator $-G_{qq}[R^\ast(N)]$ at
    distances $R^\ast(N)$ according to Eq.\ (\ref{eq:rast}) for dynamical
    triangulations of sizes ranging from $N=1000$ to $N=500\,000$ in a log-log
    plot. The solid line shows a fit of the functional form (\ref{eq:lhalfscaling})
    to the data.}
  \label{fig:lhalfscaling}
\end{figure}

For the case of the quantum gravity graphs, we determine the asymptotic behavior of
the correlator (\ref{eq:fract_av}) by fitting a suitably parametrized function to the
numerical data. For the algebraic decay expected for this graph type, we make the
ansatz,
\begin{equation}
  \label{eq:decay}
  G_{qq}(R) = A R^{-a}(1+BR^b),
\end{equation}
for $R\ge 1$, taking into account an additional effective scaling correction with
exponent $b$. From the discussion in App.\ \ref{sec:fss_corr} we expect $b=d_h=4$ to
be a reasonable choice as long as no additional, more relevant, non-analytic
corrections are present\footnote{An implicit assumption in applying such scaling
  arguments is that the ensemble of considered random graphs behaves as a system of
  statistical mechanics at a critical point. From the dynamical triangulations
  approach to quantum gravity it is known that this is indeed the case, see, e.g.,
  Ref.\ \onlinecite{ambjorn:book}}. Note that, since this ``correction'' term $R^b$
is more singular than the leading term $R^{-a}$, this form is merely an effective
description for a finite graph and distances small compared to the linear extent of
the graph. Figure \ref{fig:corr_fits} depicts the behavior of the correlator of
co-ordination numbers for the case of quantum gravity graphs as well as fits of the
form (\ref{eq:decay}) to the data. We expect the range of applicability of the form
(\ref{eq:decay}) to be limited at both sides: for very small distances $R$,
discretization effects are observed, whereas for large distances finite-size effects
modify the expected behavior, i.e., higher-order terms neglected in Eq.\ 
(\ref{eq:decay}). To account for these limitations, sampled points from both sides of
the $R$ range are successively dropped from the fit while monitoring the
goodness-of-fit parameters $\chi^2$ resp.\ $Q$.\footnote{Note that due to the
  correlations between values of $G_{qq}(R)$ for different distances $R$, the {\em
    absolute\/} values of $\chi^2$ resp.\ $Q$ are not immediately meaningful;
  relative changes, however, are.} We find restricted fits with the correction
exponent $b$ fixed at values $b=4$ or, alternatively, $b=3$, to match the data
reasonably well. A full four-parameter fit with variable exponent $b$ yields the
intermediate value $b=3.15(57)$, in total indicating the presence of higher-order
resp.\ non-analytic corrections. The fit results are compiled in Table
\ref{tab:phi3fits}.  Since we consider the correction term of Eq.\ (\ref{eq:decay})
as an effective description, we take the fit with a variable value of the correction
exponent $b$ as the most reliable and quote as our best estimate of the decay
exponent from this method
\begin{equation}
  a=1.919(76).  
\end{equation}
Due to the cross-correlations of $G_{qq}(R)$ for different distances $R$, the errors
estimated by standard fit routines are biased; thus, instead, the errors were
estimated by jackknifing over the whole fitting procedure.  As a check of whether the
used number of replicas in the disorder average is sufficient, we also performed the
same analysis with subsets of the 100 realizations.  Apart from the effect of rather
sudden jumps of the fits into different, nearby minima, which always tend to occur
with non-linear fitting procedures, we find all results completely consistent with
each other within statistical errors. Thus, e.g., for the unrestricted fit of the
form (\ref{eq:decay}) we find (using the same ranges
$R_\mathrm{min}$--$R_\mathrm{max}$ as before) $a=2.032(115)$ using only 50 graphs
and $a=1.885(216)$ using only 10 graphs.

A different way to determine the decay exponent $a$ is based on a direct application
of the finite-size scaling (FSS) behavior of the correlator. From FSS arguments it is
known that the value of the correlator at a distance $R^\ast(N)$ scaled linearly with
the size of the system behaves as
\begin{equation}
  \label{eq:lhalfscaling}
   G_{qq}[R^\ast(N)] \sim N^{-a/d_h},
\end{equation}
cf.\ App.\ \ref{sec:fss_corr}. The position of the reference points $R^\ast$ should
be selected well within the scaling region of the considered graph size. Here, we
take the largest used graph size as a reference and choose
\begin{equation}
  R^\ast(N) = 15\left(\frac{N}{5\times10^5}\right)^{1/d_h},
  \label{eq:rast}
\end{equation}
where $d_h=4$. Obviously, the resulting distances $R^\ast(N)$ will in general be
fractional numbers, for which no data are directly available. Since the accessible
{\em linear\/} graph sizes $N^{1/d_h}$ are rather small, this discretization effect
is quite pronounced. To circumvent this problem, we take a fit of the functional form
(\ref{eq:decay}) to the correlator with the correction term included as a suitable
interpolation of the data used to estimate the value of $G_{qq}[R^\ast(N)]$. As an
estimate of the statistical error, we take the error computed for the closest integer
distance. Figure \ref{fig:lhalfscaling} shows the resulting scaling plot and a fit of
the functional form (\ref{eq:lhalfscaling}) to the data. We find the data not precise
enough to resolve any corrections to the expected scaling form. The fit, including a
series of graph sizes ranging from $N=1000$ to $N=500\,000$, yields an estimate for
the decay exponent of
\begin{equation}
  a=2.09(13),  
\end{equation}
with quality $Q=0.53$. For completeness, we have also performed the same analysis for
the case of random Delaunay triangulations. As expected, the resulting values of
$G_{qq}[R^\ast(N)]$ do not show scaling according to Eq.\ (\ref{eq:lhalfscaling}),
backing up our conjecture that the correlations are not algebraic in this case.

\begin{figure}[tb]
  \centering
  \includegraphics[clip=true,keepaspectratio=true,width=\columnwidth]{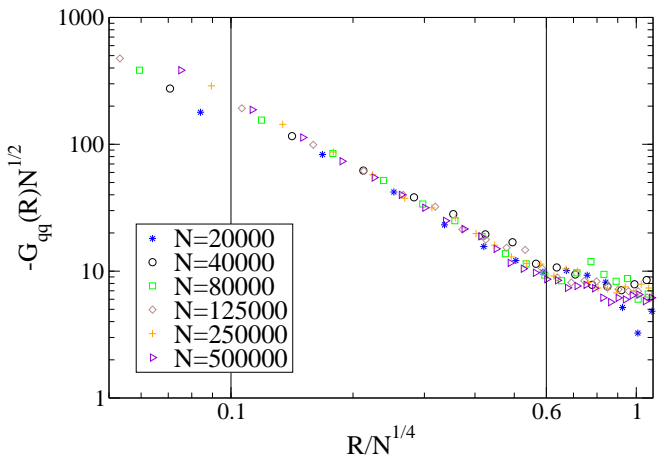}
  \caption
  {Scaling collapse on the universal scaling function $\hat{W}$ defined in Eq.\ 
    (\ref{eq:fss_correlator}) of the co-ordination number correlator $G_{qq}(R)$ for
    dynamical triangulations of sizes $N=20\,000$ to $N=500\,000$ triangles. The
    vertical lines indicate the extent of the scaling window.}
  \label{fig:scalingcollapse}
\end{figure}

Exploiting the FSS form (\ref{eq:fss_correlator}) of the correlator, it is also
possible to re-scale the data such as to model the universal FSS function $\hat{W}$
introduced in Eq.\ (\ref{eq:fss_correlator}). Thus, plotting $G_{qq}(R)N^{a/d_h}$ as
a function of the reduced distance $R/N^{1/d_h}$ should, within the scaling region,
produce data lying on a single master curve, irrespective of the lattice volume $N$
under consideration. Such a scaling plot is shown in Fig.\ \ref{fig:scalingcollapse},
where a nice scaling collapse of the data for an intermediate range of reduced
distances $R/N^{1/d_h}$ is observed. For producing this scaling plot, we assumed
$a=2$, in agreement with the results found so far and further evidence to be
presented below.  The quality of the collapse does not change visibly on slightly
changing the value of the exponent $a$ to one of the values $a=1.919$ or $a=2.09$
found before. The extension of the scaling window in reduced distance $R/N^{1/d_h}$
can be very nicely read off from Fig.\ \ref{fig:scalingcollapse} to be
\begin{equation}
  0.1 \lesssim \frac{R}{N^{1/d_h}} \lesssim 0.6,
\end{equation}
yielding, e.g., $1\le R\le 7$ for $N=20\,000$ and $3\le R\le 16$ for $N=500\,000$, in
perfect agreement with the observations from the direct analysis of the correlator.

Extending the analogy between the magnetic correlation function in a critical spin
system and the correlator considered here, one notes that the decay exponent $a$
defined above is related to the conventional critical exponents by $a=2-d_h-\eta$. A
convenient and very precise method of determining $\eta$ is to consider the scaling
of the integrated correlation function corresponding to the susceptibility instead of
analyzing the correlation function directly. The associated critical exponent
$\gamma/\nu$ then allows one to infer $\eta$ via the scaling relation $\eta =
2-\gamma/\nu$. Unfortunately, however, it follows from the relation (\ref{euler})
that the average co-ordination number on a closed triangulation is a fixed number,
depending only on the number of triangles $N$. Thus the corresponding susceptibility
always vanishes identically, such that the method outlined above cannot be applied
here.

\subsection{Averaged Fluctuations}
\label{sec:average_fluct}

\begin{figure}[tb]
  \includegraphics[width=\linewidth,clip=true,keepaspectratio=true]{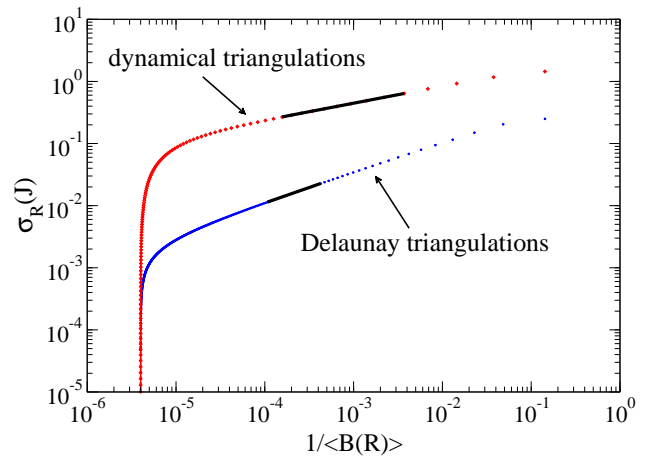}
  \caption{Decay of the averaged fluctuation $\sigma_R(J)$ of co-ordination numbers
    against the inverse averaged integrated shell volume $1/\l B(R)\r$ according to
    Eqs.\ (\ref{mean_J}) and (\ref{fluct}) for dynamical triangulations (upper
    points) and Poissonian Delaunay triangulations (lower points). For both ensembles
    lattices with $N=500\,000$ triangles have been used. The solid lines show fits of
    the functional form (\ref{fluct}) to the data.}
  \label{fig:luck}
\end{figure}

Instead of an analysis of the correlator of co-ordination numbers, the wandering
exponent $\omega$ can be directly estimated by considering the scaling of the
averaged fluctuations of co-ordination numbers and recurring to relation
(\ref{fluct}). This approach has the advantage of giving a numerical result also for
the case of correlations decaying other than algebraically, as we saw for the
Poissonian Vorono\"{\i}-Delaunay random lattices. We define the average fluctuation
$J(R)$ as indicated in Eq.\ (\ref{mean_J}) with the average performed on the level of
the fraction, in complete analogy with the definition (\ref{eq:fract_av}) of the
correlator. As before, the quenched average is performed over one starting vertex per
1000 vertices of the graph as well as over 100 different graph realizations for each
graph size. Figure \ref{fig:luck} shows the resulting decay of $\sigma_R(J)$ against
the inverse averaged integrated shell volume $1/\l B(R)\r$. The smaller number of
data points for the case of dynamical triangulations results from the smaller
effective linear extent $\sim N^{1/d_h}$ due to their large fractal dimension. The
relation of the two data sets nicely illustrates the much stronger correlations
present in the dynamical triangulations model. From Eq.\ (\ref{fluct}), we expect a
linear decline of the curve in a logarithmic presentation, the slope being given by
$1-\omega$. On the other hand, corrections to the conjectured scaling behavior for
very small distances $R$ due to discretization effects as well as for large distances
comparable to the effective linear extent of the lattices have to be taken into
account. In Fig.\ \ref{fig:luck} the scaling window of algebraic behavior according
to Eq.\ (\ref{fluct}) is nicely visible. The dramatic, exponential drop of the
fluctuations as $1/\l B(R)\r$ approaches\footnote{The number of vertices of a
  spherical triangulation consisting of $N$ triangles is $N_0=2+N/2$.} $1/(2+N/2)$ is
an effect of the topological constraint (\ref{euler}), as a consequence of which the
fluctuation $\sigma_R(J)$ vanishes identically at the maximum observed distance
$R=R_\mathrm{max}$. To obtain reliable estimates for the wandering exponent $\omega$
from a fit of the functional form (\ref{fluct}) to the data, we again successively
drop points from either side of the interval of distances $R$ while monitoring the
goodness-of-fit parameters $\chi^2$ resp.\ $Q$.\footnotemark[84] Note that, as
before, due to the cross-correlations between the values of $\sigma_R(J)$ for
neighboring distances $R$, we have to resort to jackknifing over the whole fitting
procedure to arrive at reliable error estimates.

\begin{figure}[tb]
  \includegraphics[width=\linewidth,clip=true,keepaspectratio=true]{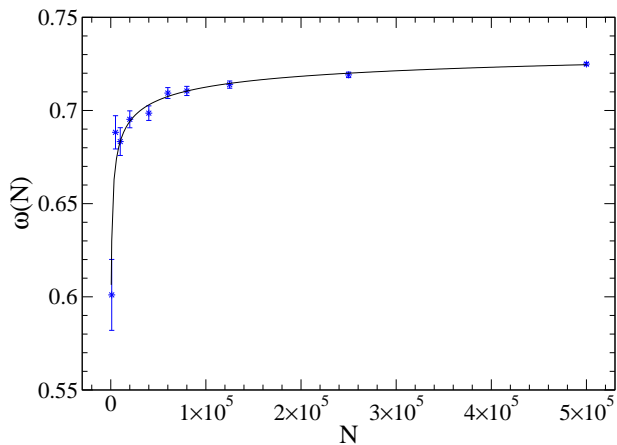}
  \caption{Finite-size scaling of the estimates of the wandering exponent $\omega(N)$
    from fits of the functional form (\ref{fluct}) to the data for dynamical
    triangulations of sizes $N=1000$ up to $N=500\,000$ triangles. The line shows a
    fit of the form (\ref{eq:fss_aproach}) to the data.}
  \label{fig:omega_fss}
\end{figure}

From the fits to the data for the maximum graph sizes $N=500\,000$ depicted in Fig.\
\ref{fig:luck}, we arrive at the following estimates for $\omega$,
\begin{equation}
  \omega = \left\{
      \begin{array}{lll}
        0.50096(55), & R=21,\ldots,41, & \mbox{Delaunay tr.}, \\
        0.72492(86), & R=5,\ldots,14,  & \mbox{dynamical tr.}. \\
      \end{array}
    \right.
  \label{eq:omegamax}
\end{equation}
From the experience with the analysis of the correlator presented above in Sec.\ 
\ref{sec:correlators}, we expect rather pronounced finite-size effects still to be
present at least for the case of the dynamical triangulations, such that the quoted
statistical error certainly does not account for all of the deviation from the
asymptotic result. Note that the result for Delaunay triangulations is perfectly
consistent with a wandering exponent $\omega=1/2$ resulting from either power-law
correlations with an exponent $a>2$, which, however, did not show up in the analysis
of the correlator presented above, or an exponential decline of correlations in
agreement with the observed non-scaling of the correlator. To account for the
suspected additional finite-size corrections present in the estimate
(\ref{eq:omegamax}) for dynamical triangulations, an additional FSS analysis is
conducted by performing the analysis described above for the series of different
graph sizes under consideration, ranging from $N=1000$ to $N=500\,000$. The resulting
FSS plot is depicted in Fig.\ \ref{fig:omega_fss}. We expect a scaling approach of
the following form,
\begin{equation}
  \omega(N) = \omega_\infty + AN^{-\theta},
  \label{eq:fss_aproach}
\end{equation}
with an {\em a priori\/} unknown correction exponent $\theta$. A fit of this form to
the data yields:
\begin{equation}
  \begin{array}{rcl}
    \omega_\infty & = & 0.7473(98), \\
    A & = & -0.73(37), \\
    \theta & = & 0.264(70),\\
    Q & = & 0.28,\\
  \end{array}
\end{equation}
Note that the correction exponent is close to $1/d_h=1/4$, which would correspond to
an analytic scaling correction. Fixing $\theta=1/4$, we find the very similar result
$\omega_\infty=0.7493(23)$. For the Poissonian Vorono\"{\i}-Delaunay triangulations,
on the other hand, we find only pretty small variations of the estimates $\omega(N)$
with the graph size, which are about of the same size as the statistical
fluctuations. Especially, for the largest two or three graph sizes, there is no
visible drift between the estimates, such that we can safely take the result
(\ref{eq:omegamax}) for $N=500\,000$ triangles as our final estimate for the
wandering exponent there.

\begin{table}[tb]
  \caption{Decay exponent $a$ resp.\ wandering exponent $\omega$ related by Eq.\
    (\ref{eq:omega_a}) for the ensemble of
    regular dynamical triangulations as estimated by various scaling methods. In each
    row, the directly measured value is printed in bold face.}
  \begin{center}
    \begin{tabular*}{\linewidth}{@{\extracolsep{\fill}}cr@{\extracolsep{0mm}.}l@{\extracolsep{\fill}}r@{\extracolsep{0mm}.}l@{\extracolsep{\fill}}}
      \toprule
      Method & \multicolumn{2}{c}{$a$} & \multicolumn{2}{c}{$\omega$} \\ \colrule
      Eq.\ (\ref{eq:decay}) & \bf 1&\bf 919(76) & 0&7601(95) \\
      Eq.\ (\ref{eq:lhalfscaling}) & \bf 2&\bf09(13) & 0&739(16) \\
      Eq.\ (\ref{eq:fss_aproach}) & 2&021(78) & \bf 0&\bf 7473(98) \\ \colrule
      average & 1&987(76) & 0&7516(95) \\ \botrule
      \label{tab:final}
    \end{tabular*}
  \end{center}
\end{table}

Finally, we also checked for the influence of the number of disorder replicas on the
results for the wandering exponent from the averaged fluctuations. For the Delaunay
triangulations of size $N=500\,000$, using 50 resp.\ 10 instead of 100 realizations
yields estimates of $\omega=0.50046(73)$ resp.\ $\omega=0.49818(469)$, in very good
agreement with the result for the full number of replicas. In the case of the
dynamical triangulations graphs, reducing the number of replicas to 50 resp.\ 10 and
performing the fits (\ref{eq:fss_aproach}) with the correction exponent $\theta$
fixed at $1/4$, we arrive at $\omega=0.7481(31)$ resp.\ $\omega=0.7590(72)$. With an
unconstrained correction exponent $\theta$, we find the data with only 10 replica not
precise enough to reliably apply the non-linear fit procedure. In Table
\ref{tab:final} we collect the final results for the decay exponent $a$ resp.\ the
wandering exponent $\omega$ for the case of dynamical triangulations from the various
methods applied, using Eq.\ (\ref{eq:omega_a}) with $d_h=4$ to compute $\omega$ from
the directly estimated decay exponent $a$ or vice versa.

\section{Conclusions}
\label{sec:conclusions}

We have considered the applicability of a relevance criterion of the Harris-Luck type
to the problem of coupling lattice models of statistical mechanics to the ensembles
of Poissonian Vorono\"{\i}-Delaunay random lattices and dynamical triangulations
resp.\ planar, ``fat'' $\phi^3$ Feynman diagrams. Following Luck's extension of
Harris' original argument to the case of systems with aperiodicity, a relevance
criterion is formulated for the case of random graphs with connectivity disorder,
i.e., a random distribution of co-ordination numbers. Depending on a characteristic
of the spatial correlations of the disorder degrees of freedom termed the wandering
exponent $\omega$, the threshold of relevance for a model with specific-heat exponent
$\alpha$ predicted by the relevance criterion is shifted from Harris' value
$\alpha_c=0$ to somewhere in the interval $-\infty<\alpha_c\le 1$, depending on the
value of $\omega$.

To determine the values of the wandering exponent for the considered ensembles of
random graphs, we have employed a detailed series of scaling studies. First, we
directly considered the behavior of the connected two-point correlation function of
the co-ordination numbers of the vertices of the triangulations. For the Poissonian
random lattices we find the correlations to decay very rapidly, and our analysis
indicates that this decline is faster than any power of the distance. For dynamical
triangulations, on the other hand, we find much stronger correlations, exhibiting a
power-law decay within the scaling window.  Due to the large fractal dimension of
these lattices, however, we find quite strong corrections to the leading scaling
behavior to be present. Taking the correction terms carefully into account, we are
able to determine the decay exponent $a$ consistently with different methods, namely
a direct analysis of the correlator, a finite-size scaling study of the correlation
function at fixed distances and a scaling collapse of the data on a universal scaling
function. Via Eq.\ (\ref{eq:omega_a}) this yields then an estimate for the wandering
exponent $\omega$.  In addition to this analysis of the correlator, we investigate
the scaling of the averaged fluctuations in spherical patches of increasing size, an
expression which directly occurs in the definition of the wandering exponent. For the
Vorono\"{\i}-Delaunay random lattices we find an estimate for the wandering exponent
consistent with a value of $\omega=1/2$ to high precision, which is in turn in
agreement with the conjecture of correlations decaying faster than any power of the
distance. Thus from Eq.\ (\ref{luckother}), for Delaunay triangulations the presented
relevance criterion reduces to the Harris criterion $\alpha_c=0$, such that disorder
of this type should be relevant for any model with positive specific-heat exponent.
For dynamical triangulations, on the other hand, a combination of the analyses
presented allows us to conjecture\footnote{Note that all critical exponents of the
  dynamical triangulations model, which could be calculated exactly, where found to
  be rational numbers, see, e.g., Ref.\ \onlinecite{ambjorn:book}.}  that the
correlator decays with an exponent $a=2$, leading to a value of the wandering
exponent of $\omega=3/4$.  In view of the relevance criterion (\ref{luckother}), this
leads to a relevance threshold as small as $\alpha_c=-2$. As a consequence, the class
of dynamical triangulations graphs should be a relevant perturbation for all known
ordered models of statistical mechanics.

There has been a number of numerical and analytical investigations of the effect of
disorder from the classes of graphs considered here on the two-dimensional Potts
model. For the case of a quenched ensemble of dynamical triangulations, a series of
Monte Carlo simulations of the $q=2$, $3$, $4$ Potts model with $\alpha=0$, $1/3$,
$2/3$, respectively, showed a change of their critical exponents, indicating a shift
to new universality classes \cite{wj:00a,wj:00b,wernecke:prep}. Additionally, it
appears that the first-order transitions of the Potts models with $q>4$ get softened
to second-order ones as the model is coupled to a quenched ensemble of dynamical
triangulations \cite{wj:00a}. For the case of percolation, corresponding to the
$q\rightarrow 1$ limit of the Potts model, for which due to its non-interacting
character quenched and annealed averages coincide, an exact solution can be obtained
by use of matrix model methods \cite{kazakov:89a}.  This solution as well as
numerical simulations for this model \cite{harris:94b}, which has $\alpha=-2/3$, also
yield changed critical exponents, in agreement with the relevance threshold obtained
here. For the case of Poissonian random lattices, simulations of the Ising or $q=2$
Potts model, corresponding to the marginal case $\alpha=0$, yielded unchanged Onsager
values for the critical exponents \cite{espriu:86a,wj:93b,wj:94a,lima:00b}; similar
results where obtained for percolation \cite{hsu:01a}, both results being in
agreement with the relevance threshold $\alpha_c=0$. On the other hand, for the
$q=3$, $4$ Potts models with $\alpha>0$, Poissonian random lattices should be a
relevant perturbation. However, an exploratory Monte Carlo study \cite{lima:00a} as
well as preliminary results from a high-precision series of Monte Carlo simulations
of the authors \cite{prep} for the three-states Potts model do {\em not\/} show any
change of universal behavior, in contradiction with the relevance threshold proposed
here. It remains an open question to be answered, e.g., by further high-precision
analyses of the $q=3$ case as well as the larger-$\alpha$ $q=4$ Potts and Baxter-Wu
\cite{baxter:book} models, whether these findings are merely the effect of a
crossover to new universal behavior occurring only for extremely large lattice sizes
or whether there is possibly some physical reason for an argument of the Harris-Luck
type not being applicable to the case of spin models coupled to Poissonian random
lattices. For this purpose, a careful analysis of the counter-examples found in
Refs.\ 
\onlinecite{derrida:85a,mukherji:95a,magalhaes:98a,haddad:00a,efrat:01a,pazmandi:97a,aharony:98a,korzhenevskii:98a}
might be instructive.












\begin{acknowledgments}
  This work was partially supported by the EC research network HPRN-CT-1999-00161
  ``Discrete Random Geometries: from solid state physics to quantum gravity'' and by
  the German-Israel-Foundation (GIF) under contract No.\ I-653-181.14/1999. M.W.\ 
  acknowledges support by the DFG through the Graduiertenkolleg ``Quantenfeldtheorie''.
\end{acknowledgments}

\appendix

\section{Nearest-Neighbor Correlations for Quantum Gravity Graphs}
\label{sec:near-neighb-corr}

This Appendix is devoted to a short derivation of the result for the co-ordination
number correlator at distance one, $G_{qq}(1)$, for the ensemble of quantum gravity
graphs mentioned in Sec.\ \ref{sec:correlators}. Godr\`eche {\em et al.\/}
\cite{godreche:92a} consider topological correlations in the thermodynamic limit of
the ensemble of (regular) dynamical triangulations via a generating-function
technique. This allows them to compute the probability distribution $Q_{ln}$ of
finding an edge connecting a vertex with co-ordination number $l$ with a vertex with
co-ordination number $n$ by means of a series expansion. This quantity is related to
the probability of finding an $l$-vertex in the neighborhood of an $n$-vertex, which
is of interest here, as follows:
\begin{equation}
  P_{ln} = \frac{1}{Z}\frac{36}{ln}Q_{ln},
  \label{eq:qlnp_rel}
\end{equation}
where
\begin{equation}
  Z=\sum_{l,n}\frac{36}{ln}Q_{ln}.
  \label{eq:z}
\end{equation}
Then, the distance-1 correlator can be expressed as
\begin{equation}
  G_{qq}(1) = \sum_{l,n} P_{ln}(6-l)(6-n) = \frac{36}{Z}(1-12S+Z),
\end{equation}
where $S=\sum_{l,n}Q_{ln}/l$. Integrating the generating function of $Q_{ln}$,
\begin{equation}
  Q(x,y) = \sum_{l,n}Q_{ln}x^ly^n,
\end{equation}
one arrives at:
\begin{equation}
  \begin{array}{rcl}
    P(y) & = & \ds\int_0^1\!\d x\,\frac{Q(x,y)-Q(0,y)}{x} =
    \sum_{l,n}\frac{Q_{ln}}{l}y^n, \\
    Z & = & 36 \ds\int_0^1\!\d y\,\frac{P(y)-P(0)}{y}=36 \sum_{l,n}\frac{Q_{ln}}{ln}.
    \label{eq:doubleintegral}
  \end{array}
\end{equation}
The first integral can be performed exactly, yielding $S=P(1)=1/6$. The double
integral in the second line of Eq.\ (\ref{eq:doubleintegral}) could not be evaluated
in closed form.  Instead, the series expansion of $Q(x,y)$ used in Ref.\ 
\onlinecite{godreche:92a}, performed up to orders $l\le 50$ and $n\le 100$ (and
additionally exploiting the symmetry property $Q_{ln}=Q_{nl}$) yields $Z\approx
0.96697$. Hence, we arrive at
\begin{equation}
  G_{qq}(1) = 36\frac{Z-1}{Z} \approx -1.2295.
\end{equation}

\section{Finite-Size Scaling of the Correlator}
\label{sec:fss_corr}

In this Appendix we give a short justification of the finite-size scaling (FSS)
method used in Sec.\ \ref{sec:correlators} to determine the scaling exponent of the
two-point correlator of co-ordination numbers and the correction terms taken into
account. We consider making a real-space renormalization transformation with a
re-scaling factor $b$ (see, e.g., Refs.\ 
\onlinecite{barber:domb,cardy:book,henkel:book}). After $n$ iterations of the
re-scaling, the two-point correlation function of some operator $\phi$ under
consideration can be written as
\begin{equation}
  G_{\phi\phi}(R;t,{\ts\frac{1}{N}}) = b^{-2x_\phi n}\,G_{\phi\phi}\left[\frac{R}{b^n};tb^{n
  y_t},\frac{1}{N}b^{n d_h}\right],
\end{equation}
where $x_\phi$ denotes the scaling dimension of the operator $\phi$ and $y_t$ the
temperature-related renormalization-group eigenvalue. Stopping the scale
transformation at an iteration such that $N^{-1}b^{nd_h}\equiv K$, i.e.,
$b^n=(KN)^{1/d_h}\equiv (N/N_0)^{1/d_h}$, we arrive at
\begin{equation}
  G_{\phi\phi}(R;0,{\ts\frac{1}{N}}) = \left(\frac{N}{N_0}\right)^{-\frac{2x_\phi}{d_h}}
  \hat{W}\left[\frac{R}{(N/N_0)^\frac{1}{d_h}}\right],
  \label{eq:fss_correlator}
\end{equation}
introducing a universal scaling function $\hat{W}$. Therefore, if $R$ is scaled
linearly with $N^{1/d_h}$, e.g., $R^\ast=N^{1/d_h}/2$, the correlation function
scales as
\begin{equation}
  G_{\phi\phi}(R^\ast;0,{\ts\frac{1}{N}}) \sim N^{-\frac{2x_\phi}{d_h}}.
\end{equation}

Now, instead of doing FSS, consider the finite-size ``field'' as a scaling
correction and stop the re-scaling at $R/b^n=K$,
\begin{equation}
  G_{\phi\phi}(R;t,{\ts\frac{1}{N}})=\left(\frac{R}{R_0}\right)^{-2x_\phi}
  F\left[t\left(\frac{R}{R_0}\right)^{y_t},
    \frac{1}{N}\left(\frac{R}{R_0}\right)^{d_h}\right],
\end{equation}
where $F$ is another scaling function. Hence, at criticality one has
\begin{equation}
  \begin{array}{rcl}
    G_{\phi\phi}(R;0,\frac{1}{N}) & = & \ds\left(\frac{R}{R_0}\right)^{-2x_\phi}
    \hat{F}\left[\frac{1}{N}\left(\frac{R}{R_0}\right)^{d_h}\right] \\
    & \approx & \ds\left(\frac{R}{R_0}\right)^{-2x_\phi}
    \left(\hat{F}(0)+\frac{1}{N}\left(\frac{R}{R_0}\right)^{d_h}
      \hat{F}'(0)\right),
  \end{array}
\end{equation}
which should be a reasonable approximation as long as $R\ll N^{1/d_h}$.


\end{document}